# A Revisit to High Thermoelectric Performance of Single-layer MoS$_2$

Zelin Jin[1, #], Quanwen Liao[1, #], Haisheng Fang[1,*], Zhichun Liu[1,*], Wei Liu[1], Zhidong Ding[1], Tengfei Luo[3], Nuo Yang[1,2,*]

[1] School of Energy and Power Engineering, Huazhong University of Science and Technology (HUST), Wuhan 430074, People's Republic of China

[2] State Key Laboratory of Coal Combustion, Huazhong University of Science and Technology (HUST), Wuhan 430074, China

[3] Department of Aerospace and Mechanical Engineering, University of Notre Dame, Notre Dame, Indiana 46556, USA.

[#] Z. J. and Q. L. contributed equally to this work.

* Corresponding authors: N.Y. (nuo@hust.edu.cn), H. F. (hafang@hust.edu.cn), and Z.L.(zcliu@hust.edu.cn)

**Abstract**

Both electron and phonon transport properties of single layer $MoS_2$ ($SLMoS_2$) are studied. Based on first-principles calculations, the electrical conductivity of $SLMoS_2$ is calculated by Boltzmann equations. The thermal conductivity of $SLMoS_2$ is calculated to be as high as 116.8 $Wm^{-1}K^{-1}$ by molecular dynamics (MD) simulations. The predicted value of *ZT* is as high as 0.26 at 500K. As the thermal conductivity could be reduced largely by phonon engineering, there should be a high possibility to enhance *ZT* in the $SLMoS_2$-based materials.

**Introduction**

Thermoelectric materials are essential for converting waste heat to electricity and solid-state cooling, which have attracted much attention recently [1-10]. The dimensionless figure of merit (*ZT*) is utilized to evaluate the efficiency of the thermoelectric conversion as: $ZT = S^2\sigma T/\kappa$, where $S$ is the Seebeck coefficient, $\sigma$ is the electrical conductivity, $T$ is the absolute temperature, and $\kappa$ is the total thermal conductivity. The $\kappa$ is composed of electrons' contribution ($\kappa_e$) and phonons' contribution ($\kappa_p$). The *ZT* value for most commercial materials are around one, which is far below the critical value of three that can revolutionize the traditional energy conversion [4]. In the past two decades, nano- and nano-structured- materials are expected to have excellent energy conversion efficiency due to the higher power factor ($S^2\sigma$) [11, 12] and lower $\kappa_p$ [13-15], which are also known as the electron-crystal and phonon-glass.

Graphene, as the first two dimensional material, has extraordinary electronic property as well as super high thermal conductivity [16]. However, the pristine graphene, a semi-metal, has zero band gap and very small $S$ [17]. Different from graphene, single layer MoS$_2$ (SLMoS$_2$) is a semiconductor and has a direct band-gap [18], which enables its wide applications in electronic and optical devices, such as field effect transistor [19].

Recently, some works have studied the electronic and phononic properties of SLMoS$_2$. Eugene *et al.* have calculated the electronic structure of SLMoS$_2$ and compared the results to bulk MoS$_2$ [20], and revealed the transition mechanism of band gap from indirect (SLMoS$_2$) to direct (bulk MoS$_2$). Emilio *et al.* have shown that, after applying compressive or tensile bi-axial strain, the electronic structure of SLMoS$_2$ transitions from semiconductor to metal [21]. Li *et al.* calculated the intrinsic electrical transport and electron-phonon interaction properties of SLMoS$_2$ [22]. Moreover, the thermoelectric potential of SLMoS$_2$ has been explored and a maximum *ZT*, at room temperature, is obtained as 0.5 by Huang *et al.* [23] using the ballistic model. The scatterings of electrons are not considered in their ballistic model, which should have led to an over-estimation of *ZT*. Fu *et al.* studied SLMoS$_2$ ribbons and calculated the *ZT* value to be up to 3.4 [24]. Besides theoretical predictions, Wu *et al.* has experimentally reported a large *S* of 30 mV/K for SLMoS$_2$ [25], which indicates an appealing potential for thermoelectric applications.

Besides electron properties, some works focused on the phonon properties of SLMoS$_2$ [26]. SLMoS$_2$ nanoribbon has a low thermal conductivity due to the size effect. Jiang *et al.* claimed that $\kappa_p$ of SLMoS$_2$ nanoribbon was around 5 Wm$^{-1}$K$^{-1}$ at room temperature by molecular dynamics (MD) simulations [27]. Zhang *et al.* reported three reslutes for SLMoS$_2$ nanoribbons which are 1.35 Wm$^{-1}$K$^{-1}$ from MD [28], 23.2 Wm$^{-1}$K$^{-1}$ from nonequilibrium Green's function [29], and 26.2 Wm$^{-1}$K$^{-1}$ from Boltzmann transport equation [30]. However, there are much higher thermal conductivities for MoS$_2$ from previous reports. Li *et al.* predicts the $\kappa$ as 83 Wm$^{-1}$K$^{-1}$ from *ab initio* calculations [31]. With high-quality sample, the $\kappa$ of suspended few layers MoS$_2$ has been measured as 52 Wm$^{-1}$K$^{-1}$ [32] and 35 Wm$^{-1}$K$^{-1}$ [44]. Liu *et al.* claimed that the basal-plane thermal conductivity of single crystal MoS$_2$ would be 85-110 Wm$^{-1}$K$^{-1}$ [33]. There is not an agreement on the $\kappa$ of SLMoS$_2$, and it needs more works on this issue.

In this paper, both electron and phonon transport properties of SLMoS$_2$ are studied. Based on the electronic band structure from first-principles calculations, the electrical conductivity of SLMoS$_2$ is calculated by Boltzmann equations. Both the electronic structure and phonon dispersion relation are calculated. Together with $\kappa_p$ calculated from classical equilibrium molecular dynamics (EMD) simulations, the thermoelectric properties are obtained. The results show that SLMoS$_2$ is a promising material for thermoelectric engineering.

**Methods**

To calculate electronic properties, the first-principles calculation is implemented by QUANTUM ESPRESSO in the frame of density functional theory (DFT) [26]. The local density approximation (LDA) is used in the exchange-correlation approximation while the semi-core valence for molybdenum is considered with the Goedecker-Hartwigsen-Hutter-Tetter method [34]. The wave-functions in electronic calculation are cut off at 160 Ry, and the irreducible Brillouin zone is sampled with a $16 \times 16 \times 1$ Monkhorst-Pack grid.

Structure relaxation for SLMoS$_2$ yields lattice constant of about 3.13 Å, consistent with previous predications of 3.12-3.16 Å [20-22]. For the consistency of property evaluation, the thickness of SLMoS$_2$ is assumed to be 6.16 Å -- the same as that of the single-sheet in bulk MoS$_2$ [35]. The following calculations on electrons and phonons are based on this optimized structure.

In the calculations of transport coefficients, a k-point mesh as $28 \times 28 \times 1$ (denser enough to obtain converged results) is used over the irreducible Brillouin zone. With the assumption of constant relaxation time, the transport coefficient for electrons can be calculated using BoltzTrap [36] which solves Boltzmann transport equation (more details in supplemental information).

The thermal conductivity of phonon of SLMoS$_2$, $\kappa_p$, is calculated by EMD with the Green-Kubo approach [37]. All the simulations are carried out utilizing the LAMMPS software package [38]. The Stillinger-Weber potential with parameters fitted by Jiang *et al.* [27] is adopted in our simulations. The SLMoS$_2$ film is constructed by periodic arrangement of supercell illustrated in Fig. 1, and the sizes of 1 × 1 units$^2$ supercell are 10.826 × 9.375 nm$^2$. To study the finite size effect on thermal conductivities, we calculated the simulation cells with the volumes from 2 × 2 to 32 × 32 units$^2$ at room temperature. (more details in supplemental information).

**Results and discussions**

The electronic band structure of SLMoS$_2$ along the high-symmetry points in Brillouin zone is shown in Fig. 2(a). At the K point, there is a direct band gap as 1.86 eV which agrees well with previous calculations (1.69 ~ 1.98 eV) [20-23, 39, 40]. Another characteristic in the SLMoS$_2$ band structure is that there is a Q valley along the Γ-K path. The Q valley yields a larger effective mass than the K valley, which leads to strong electron-phonon interactions in MoS$_2$ at this point [22]. The large effective mass of carriers and multi-valleys band structure are favorable for a high *ZT* [41]. Due to the quantum size effects in 2D structrue, the density of state (DOS) electrons shows that there are sharp gradients at the edges of both conduction and valence band and several peaks near band edges, which may enhance ZT as the prediction of Mahan and Sofo [12].

Figure 3 shows a full calculation of the thermoelectric properties of SLMoS$_2$ at 300 K, 400 K and 500 K. As shown in Fig. 3(a) and (b), $\kappa_e$ and $\sigma$ increase as the increasing of carrier concentration ($n_e$). When the Fermi level is in the band gap, $n_e$ and $\sigma$ is much smaller. As the Fermi level moves up into the conduction band ($n_e > 3 \times 10^{19}$ cm$^{-3}$), the electrical conductivity increases quickly. Shown in Fig. 3(a), the seebeck coefficient is large and decreases with the increase of $n_e$. The Fermi level for ZT peak locates around the first DOS peak, and this is consistent with the prediction that a delta DOS would result in the optimum ZT.

The phonon dispersion relation of SLMoS$_2$ is also calculated and shown in Fig. 2(b). For the three lowest branches in the vicinity of Γ, there are quadratic relation for the out-of-plane transverse acoustic (ZA) phonon and linear relation for transversal acoustic (TA) and longitudinal acoustic (LA) phonon. The group velocities at Γ along Γ-M direction are around 667.5 m/s (TA) and 1080.2 m/s (LA), which are much smaller than group velocities in graphene [42], 3743 m/s (TA) and 5953 m/s (LA).

For semiconductors, the thermal conductivity is mainly contributed by phonons ($\kappa_p$). We calculated $\kappa_p$ by EMD and show in Fig. 4. The $\kappa_p$ of SLMoS$_2$ exhibits a size effect and reaches a converged value when the simulation cell is larger than 8 × 8 units$^2$ (86.6 × 75 nm$^2$) as shown in Fig. 4(a). A weak anisotropy is observed in thermal conductivities along armchair and zigzag direction. The average value of $\kappa_p$ along armchair and zigzag directions is 116.8 Wm$^{-1}$K$^{-1}$ for simulation cell as 32 × 32 units$^2$ (346.4 × 300 nm$^2$) at 300K. In Fig. 4(b), the $\kappa_p$ of SLMoS$_2$ decreases with the increasing temperature (79.6 Wm$^{-1}$K$^{-1}$ and 52.9 Wm$^{-1}$K$^{-1}$ at 400 K and 500 K, respectively), because there are more three phonon Umklapp scatterings for high temperature. A lower $\kappa_p$ can enhance thermoelectric properties.

We obtained a much higher $\kappa_p$ of SLMoS$_2$ than that of nanoribbons [31, 32] when the same empirical potential is used in MD simulations. That is, the size effect is obvious in the confinement of heat transfer. Our value is comparable to the result predicted from *ab initio* calculation [36], 83 Wm$^{-1}$K$^{-1}$ at 300 K. Due to the absence of impurities, defects and interlayer scatterings in MD simulations, the $\kappa$ of SLMoS$_2$ is a little higher than the measurements of bulk multilayer SLMoS$_2$ [33], 85~110 Wm$^{-1}$K$^{-1}$.

With the above calculations of electron and phonon properties, *ZT* profiles can be obtained and are shown in Fig. 3(d). There is a parabolic tendency for *ZT* in the whole carrier concentration range. The optimized *ZT* values are 0.04, 0.11, and 0.26 for 300 K, 400 K and 500 K, respectively. These values get bigger as temperature increases because of the improved power factors and reduced thermal conductivity. As mentioned above, these optimized *ZT* correspond to the situation where the Fermi level moves up to the first peak in the conduction band.

As shown in Table 1, we list some recently results on thermoelectric properties of different $SLMoS_2$ structure. The value of *ZT* is in the same order of Ref.[23] and one order smaller than that of $SLMoS_2$ ribbon. Our results on $\kappa_p$ and is much higher than others. As shown in Fig. 4(a), our $\kappa_p$ by EMD can overcome the size confinement effect and can get a value corresponding to an infinite $SLMoS_2$ sheet. Moreover, it do not need the assumption of linear relationship between $1/\kappa_p$ and $1/L$ when extrapolating $1/L \rightarrow 0$ in the calculation of non-equilibrium MD (NEMD).

Compared to nanoporous silicon analyzed by Lee [43], we get the similar *ZT* trend and magnitude of these transport values. As shown in Fig. 3(c), the power factor of $SLMoS_2$ is larger than that of nanoporous silicon. The large power factor of $SLMoS_2$ comes from a larger intrinsic $\sigma$ and a comparable *S*. It indicates that the $SLMoS_2$ has comparable electron properties as the optimized nanoporous silicon. However, due to the high $\kappa_e$ and $\kappa_p$, the $SLMoS_2$ gives a modest *ZT* value. It is also worth noting that the ZT value here is smaller than the prediction from ballistic models by Huang et al. which get a higher TE performance due to neglecting the phonon scatterings [23].

Although the predicted ZT value of SLMoS$_2$ is not over one as a thermoelectric material. It is worth noticing that bulk silicon has a ZT value as low as 0.003. However, with phonon engineering, Si-based nanomaterials may reach high ZT values, such as Si nanowires [44, 45], nanoporous Si [14, 43, 46, 47], and nanostructured Si [48]. Another inspiration example is the graphene. The high pristine thermal conductivity of graphene can be reduced largely by phonon engineering [49-51] and obtain a ZT as high as 3 [51]. The ZT of SLMoS$_2$ is much higher than that of silicon and graphene. With phonon engineering, SLMoS$_2$-based materials may be a good candidate for thermoelectric application.

**Conclusion**

The thermoelectric properties of SLMoS$_2$ are explored using theoretical calculations. Electronic structure and phonon dispersion relation are calculated using DFT calculations. Combined with molecule dynamics simulations and Boltzmann equations, thermoelectric properties are predicted as a function of carrier concentration at room temperature. With the lattice thermal conductivity as 116.8 Wm$^{-1}$K$^{-1}$, 79.6 Wm$^{-1}$K$^{-1}$, and 52.9 Wm$^{-1}$K$^{-1}$, the optimized ZT of SLMoS$_2$ is found to be of 0.04, 0.11 and 0.26 at 300 K, 400 K and 500 K, respectively. As SLMoS$_2$ has a higher *ZT* than other pristine structure, like Si and graphene, there should be a high possibility to enhance *ZT* in the SLMoS$_2$-based materials.

**Acknowledgments**

H.F. was supported in part by National Natural Science Foundation of China Grant (51476068). N.Y. was supported in part by the National Natural Science Foundation of China Grant (11204216). Z. L. was supported in part by the National Natural Science Foundation of China (Grant No. 51376069) and by the Major State Basic Research Development Program of China (Grant No. 2013CB228302). We are grateful to Lina Yang, and Jintao Lv for useful discussions. The authors thank the National Supercomputing Center in Tianjin (NSCC-TJ) for providing help in computations.

Table 1. The comparison of thermoelectric properties for different MoS$_2$ structures, including single layer (SL), few layers (FL), single layer ribbon (SLR), and bulk MoS$_2$.

| Struct.& Ref. | Method | $T$ (K) | Carrier type | $\sigma$ (Scm$^{-1}$) | $S$ (μVK$^{-1}$) | $\kappa_e$ (Wm$^{-1}$K$^{-1}$) | $\kappa_{ph}$ (Wm$^{-1}$K$^{-1}$) | $ZT$ |
|---|---|---|---|---|---|---|---|---|
| SL | DFT+BTE+MD | 300 | n<br>p | 14625<br>16957 | -110<br>72.9 | 8.94<br>11.39 | 116.8 | 0.04<br>0.02 |
| | | 500 | n<br>p | 11714<br>8853 | -161<br>150 | 9.69<br>8.40 | 52.9 | 0.26<br>0.16 |
| SL[23] | DFT+Ballistic model | 300 | n<br>p | 54<br>108 | -202<br>215 | 0.021<br>0.040 | 0.243<br>0.244 | 0.25<br>0.53 |
| SLR[24, 28] | DFT+BTE+MD | 300 | n<br>p | 7770<br>14300 | -204<br>223 | 2.89<br>5.20 | 1.02 | 2.5<br>3.4 |
| CVD SL[25] | Experiment | 300 | - | - | ≤30000 | - | - | - |
| SL FET[52] | Experiment | 300 | - | - | 400-100000 | - | - | - |
| Bulk[53] | Experiment | 90-873 | - | - | 500-700 | - | - | - |
| SL[28] | MD | 300 | - | - | - | - | 1.35 | - |
| SL[31] | DFT+BTE | 300 | - | - | - | - | 83 | - |
| SL[29] | DFT+NEGF | 300 | - | - | - | - | 23.2 | - |
| SL[30] | DFT+BTE | 300 | - | - | - | - | 26.2 | - |
| SLR[27] | NEMD | 300 | - | - | - | - | 5 | - |
| FL[32] | Experiment | 300 | - | - | - | - | 52 | - |
| SL[54] | Experiment | 300 | - | - | - | - | 35.4 | - |
| Bulk[33] | Experiment | 300 | - | - | - | - | 85-110 | - |

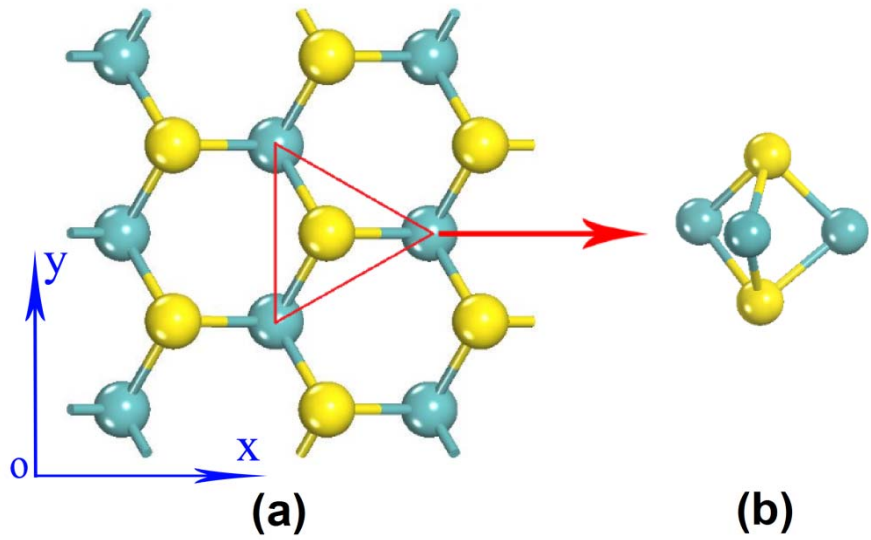

**Fig. 1** The structure of single layer MoS$_2$. (a) The top view, a hexagonal lattice structure similar to graphene. (b) The side view of the inset triangle. There are three molybdenum atoms as the first nearest neighbor of each sulfur atom. There are six sulfur atoms as the first nearest neighbor of each molybdenum atom.

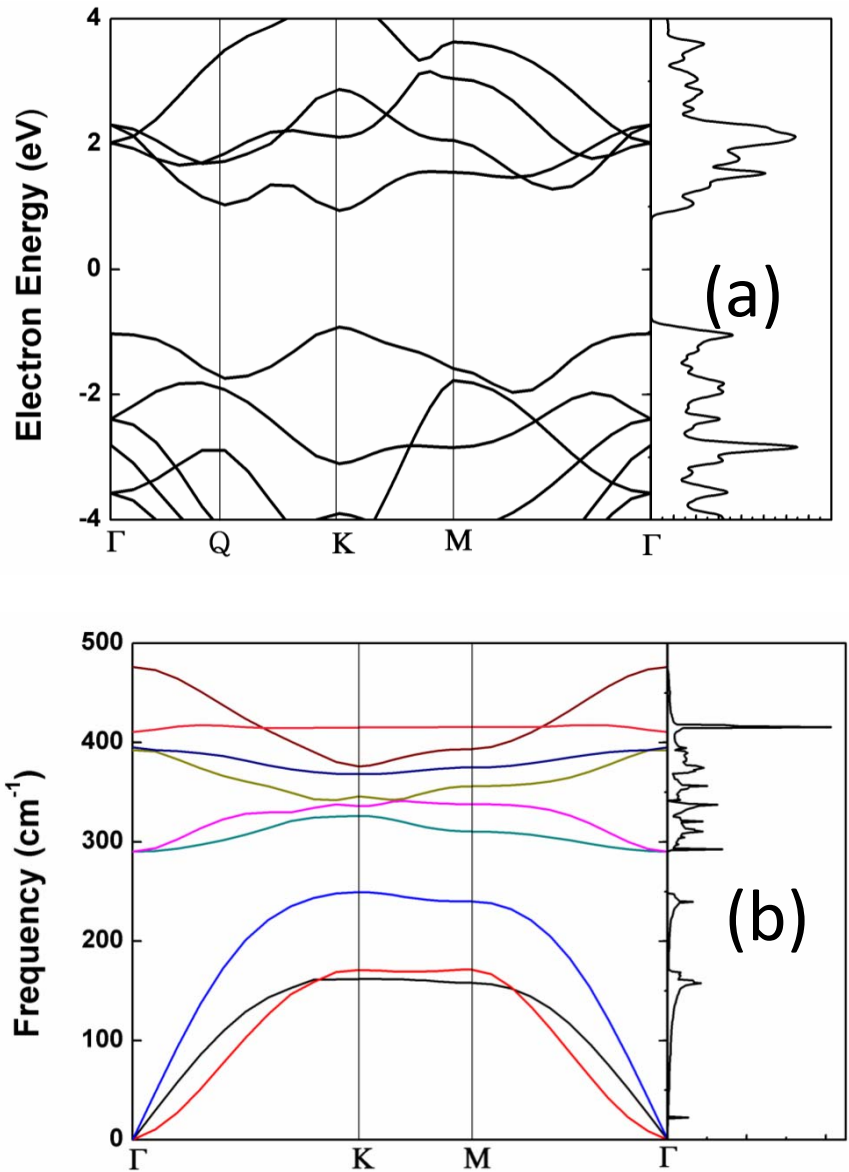

**Fig. 2** (a) Electron band structure and density of state along the symmetry line. The Fermi energy is set in the middle of the gap. (b) The phonon dispersion for SLMoS$_2$ and the phonon density of state in the whole Brillouin zone.

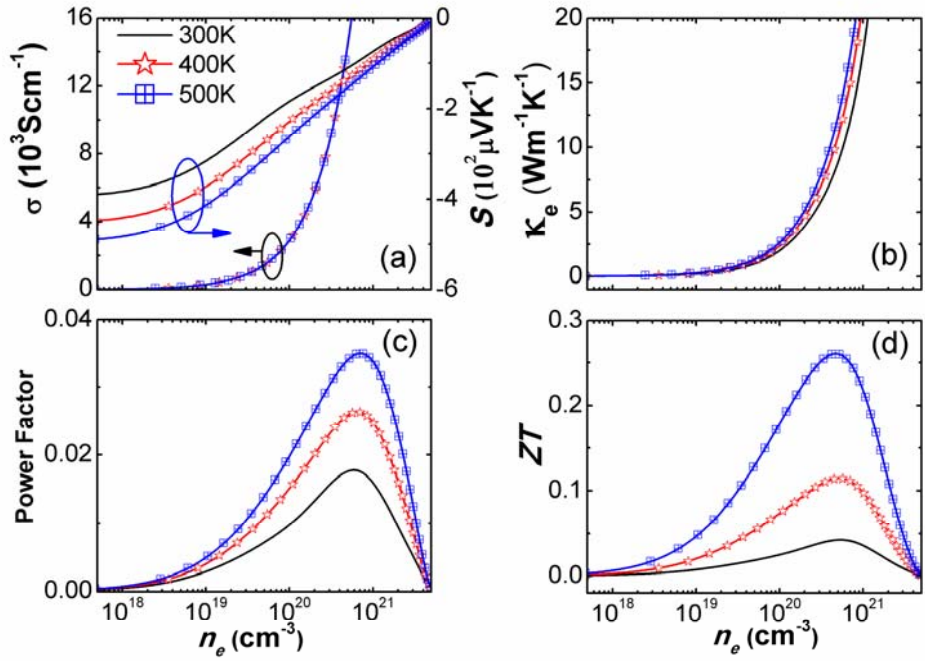

**Fig. 3** The thermoelectric transport properties of n-type SLMoS$_2$ at 300K, 400K and 500K. (a) The electrical conductivity; (b) The electrical thermal conductivity; (c) The Seebeck coefficient; (d) The figure of merit.

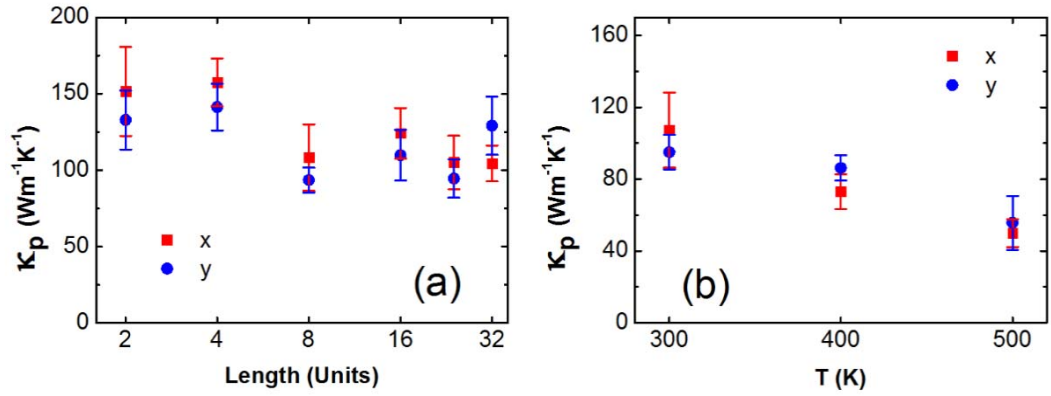

**Fig. 4** (a) The thermal conductivity dependence of SLMoS$_2$ upon size. The $\kappa_p$ is the thermal conductivity, and the length is an integer multiple of the nanoscale supercell (10.826 nm × 9.375 nm). The x is armchair direction, and the y is zigzag direction. (b) The thermal conductivity dependence of SLMoS$_2$ upon temperature. The $\kappa$ is the thermal conductivity, and the temperature is adopted as 300K, 400K and 500K, respectively. The x is armchair direction, and the y is zigzag direction.